\documentclass[pra,reprint,showpacs,superscriptaddress]{revtex4-1}
\usepackage{etex}
\usepackage{bbm}
\usepackage{amsmath}
\usepackage{amssymb}
\usepackage{epsfig}
\usepackage[utf8]{inputenc}
\newcommand{\Hid}[0]{H_{\mathrm{id}}}
\newcommand{\Henv}[0]{H_{\mathrm{env}}}
\newcommand{\Uid}[0]{U_{\mathrm{id}}}
\newcommand{\Herr}[0]{H_{\mathrm{error}}}
\newcommand{\Hint}[0]{H_{\mathrm{int}}}
\newcommand{\Havg}[0]{\overline{H}}
\newcommand{\Havgo}[0]{\Havg^{(0)}}
\newcommand{\id}[0]{\mathbbm{1}}
\newcommand{\ket}[1]{\left| #1 \right \rangle \!}
\newcommand{\bra}[1]{\left \langle #1 \right| \!}

\newcommand{\sK}[0]{\mathcal{K}}
\newcommand{\dotp}[2]{\vec{#1} \cdot \vec{#2}}

\begin{document}
\title{Constructing Pauli pulse schemes for decoupling and quantum simulation}
\author{Holger Frydrych}
\email{holger.frydrych@physik.tu-darmstadt.de}
\author{Gernot Alber}
\affiliation{Institut für Angewandte Physik, Technische Universität Darmstadt, D-64289 Darmstadt}
\author{Pavel Ba\v{z}ant}
\affiliation{Department of Physics, FNSPE, Czech Technical University in Prague,
B\v{r}ehov\'a 7, 115 19 Praha 1, Star\'e M\v{e}sto, Czech Republic}

\begin{abstract}
Dynamical decoupling is a powerful technique to suppress errors in quantum systems originating from 
environmental couplings or from unwanted inter-particle interactions. However,
it can also be used to selectively decouple specific couplings in 
a quantum system. We present a simple and easy-to-use general method to construct such
selective decoupling schemes on qubit and qudit networks by means of (generalized) Pauli operations. As these
constructed schemes can suppress Hamiltonian interactions on general qudit networks selectively, they are well suited
for purposes of approximate quantum simulation.
Some examples are presented, demonstrating the use of our method and the resulting decoupling schemes.
\end{abstract}

\pacs{03.67.Pp,03.65.Aa}

\maketitle

\section{Introduction}
An important challenge in quantum information science is to protect quantum memories and communication channels against unwanted couplings within this quantum system or to an environment. 
It is to this end that, in 1999, Viola {\it et al.} presented a promising method they called dynamical decoupling \cite{vkl99}. In this method, local control operations are used to counteract the influence of unwanted couplings with an environment. It can be seen as a generalization of techniques which were previously developed in the area of nuclear magnetic resonance. The spin-echo effect \cite{hahn50} is a well-known example of the latter. 
Over the years, dynamical decoupling has been successfully implemented in experimental setups to protect quantum systems against unwanted influences \cite{mtabplb06,fsl05,budsib09,lgdvt11,hhh13}. 

Typically, dynamical decoupling schemes assume that the quantum system to be protected 
can be manipulated by a set of fast local control operations. As a first approximation, it is often assumed that these control operations act instantaneously 
(bang-bang control \cite{vl98}) and that they can be represented by a set of unitary operators. 
By choosing an appropriate set of unitary operators and applying them in a particular order, 
it is possible to change the state of a quantum system unitarily such
that the effects of unwanted Hamiltonian interactions are averaged out approximately. 
In contrast to quantum error correction, which is capable of correcting errors perfectly, dynamical decoupling generally only suppresses unwanted couplings.

However, the use of unitary control operations is not limited to suppressing decoherence effects or unwanted intra-particle couplings within a many-particle quantum system. Viola {\it et al.} \cite{vlk99} already showed that dynamical decoupling can also be used to suppress only certain components of a many-particle Hamiltonian selectively. Thus, it is possible to develop control strategies that, within certain limits, are suitable for the purpose of quantum simulation, meaning that the effects of a present Hamiltonian interaction with active controls on a quantum system resemble those of a different Hamiltonian. 
Wocjan {\it et al.} \cite{wrjb02a} demonstrated that 
any quantum system with a non-trivial Hamiltonian can simulate any other Hamiltonian interaction, 
provided that a suitable finite set of unitary control operations is available. Dodd {\it et al.} \cite{dodd02} showed that any two-body Hamiltonian on a qubit network can be simulated by any two-body entangling Hamiltonian with the help of local unitaries, a result that was generalized to qudit systems by Nielsen {\it et al.} in \cite{nbdcd02}. 

For purposes of quantum information processing, it is of particular interest to develop error-suppressing dynamical decoupling schemes and quantum simulation schemes for systems comprised of distinguishable qubits. For these systems, a special case of decoupling controls is frequently discussed where all control operations consist of instantaneously applied Pauli operations acting locally on each qubit separately. Numerous efficient schemes of this kind have been developed which are capable of suppressing environmental errors or unwanted inter-qubit couplings in many-qubit systems. Stollsteimer and Mahler \cite{sm01}, for example, proposed a construction
based on orthogonal arrays, while Leung \cite{leu02} presented a decoupling method based on Hadamard matrices. Both approaches were eventually unified by Rötteler and Wocjan \cite{rw06}. Wocjan {\it et al.} also discussed applications of similar constructions to quantum simulation scenarios \cite{wrjb02b}.  Furthermore, several advanced control strategies have been devised to enhance the performance of basic decoupling schemes. 

Despite these interesting developments, it remains a challenge to find and implement suitable Pauli pulse schemes for the purpose of quantum simulation. While specific constructions for specific scenarios have been developed, so far there has been no systematic method for constructing dynamical decoupling schemes from simple Pauli pulses for a general scenario that applies to both error suppression and quantum simulation for arbitrary many-body Hamiltonians.

In this paper, we present a systematic method for constructing 
decoupling schemes from local Pauli pulses on networks of qubits that are capable of changing the action of a given arbitrary Hamiltonian $H$ to that of a wanted "ideal" Hamiltonian $\Hid$. Only (partial) knowledge of $H$ and $\Hid$ is required; there is no dependency on the availability of, e.g., suitable orthogonal arrays. 
This method is not only useful for protecting specific inter-particle couplings against unwanted
couplings or environmental influences, but also for simulating ideal Hamiltonian dynamics within certain limits. 
The restriction to local Pauli operations leads to a particularly simple, but still powerful procedure which exploits two basic properties of the Pauli operators, namely that they are Hermitian and unitary and that they fulfill characteristic Clifford-type algebraic relations. We will also show that certain aspects of these properties carry over to generalized spin operators, allowing our method to be generalized to qudit networks of arbitrary dimension. Although the method does not work for all possible pairs of $H$ and $\Hid$, its limitations are easily understood and still allow for a multitude of interesting applications.

The outline of this paper is as follows: 
In Sec. \ref{sec:decoupling_basics}, we briefly summarize the basics of dynamical decoupling in qubit systems which are important for our subsequent discussion. In Sec. \ref{sec:decoupling_schemes},  a general method for designing dynamical decoupling schemes based on Pauli pulses is presented which is capable of simulating an ideal many-qubit Hamiltonian $\Hid$ approximately by another many-qubit Hamiltonian $H$. In this section, it is also shown under which conditions such schemes can be developed.
In Sec. \ref{sec:examples}, our construction method is applied to some simple examples.
Finally, in Sec. \ref{sec:qudits}, the generalization of this construction to qudit networks is discussed briefly. 
Technical details concerning these higher-dimensional generalizations are presented in three appendices. A fourth appendix contains details of numerical simulations conducted to study the effectiveness of some of our schemes.

\section{Dynamical decoupling - basic facts} \label{sec:decoupling_basics}
In this section, basic known principles of dynamical decoupling on networks of qubits are summarized which are important for our subsequent discussion. 

We consider a quantum system consisting of $N$ qubits whose state space is described by the Hilbert space 
\begin{equation}
\mathcal{H} = \mathcal{H}_\text{sys} \otimes \mathcal{H}_\text{env} , \quad \mathcal{H}_\text{sys} = \bigotimes_{i=1}^N \mathbb{C}^2 , 
\end{equation}
with $\mathcal{H}_\text{env}$ the Hilbert space of an arbitrary environment. The time evolution of the whole system is governed by a Hamiltonian $H$. 

Let us assume that at the beginning of periodic time intervals of length $\Delta t$ we can apply instantaneously (bang-bang control) any of the unitary and Hermitian Pauli operators 
$\sigma_1, \sigma_2, \sigma_3$ to any qubit individually. This results in unitary pulses of the form
\begin{equation}
p_k = \sigma_{i_{k_1}} \otimes \dots \otimes \sigma_{i_{k_N}} \otimes \id_\text{env}, \quad i_{k_j} \in [0,1,2,3],
\end{equation}
where $\sigma_0 \equiv \id$ is used if no control action is taken on a particular qubit.
After $m$ such instantaneous control pulses the time evolution of the total system 
is described by the unitary operator (we assume $\hbar = 1$)
\begin{align} 
U(m \Delta t) &= p_m e^{-iH\Delta t} p_{m-1} e^{-iH\Delta t} \dots p_1 e^{-iH\Delta t} p_0 \notag \\
&= g_m (g_{m-1}^\dag e^{-iH\Delta t} g_{m-1}) \dots (g_0^\dag e^{-iH\Delta t} g_0) \notag \\
&= g_m e^{-i (g_{m-1}^\dag H g_{m-1}) \Delta t} \dots e^{-i (g_0^\dag H g_0) \Delta t},
\label{time}
\end{align}
where we introduced the operators $g_k = p_k  p_{k-1}  \dots \cdot p_0$.
This time evolution can also be described by an effective average Hamiltonian $\Havg$:
\begin{equation}
U(m\cdot \Delta t) = g_m e^{-i\Havg m\cdot \Delta t}.
\end{equation}
The operator $g_m$ is arbitrary and can be chosen to be the identity operation. 
$\Havg$ depends on $\Delta t$, and we can do a Magnus expansion \cite{magnus54} to develop $\Havg$ into a series of terms depending on increasing orders of $\Delta t$, i.e.,
\begin{equation}
\Havg = \Havgo + \Havg^{(1)} + \Havg^{(2)} + \dots .
\end{equation}

The main goal of approximate quantum simulation by dynamical decoupling is
to construct a sequence of control pulses $p_k$ such that, to the lowest order of the Magnus expansion $\Havgo$,
the original Hamiltonian $H$ is transformed into a wanted ideal Hamiltonian $\Hid$ of the $N$-qubit system which contains no couplings between the $N$ qubits and the environment and which may have modified inner couplings between the qubits.
The lowest order of the Magnus expansion is given by
\begin{equation}
\Havgo = \frac{1}{m} \sum_{i=0}^{m-1} g_i^\dag H g_i . 
\end{equation}
We call a set of operators $\{g_i\}_{i=0}^{m-1}$ a decoupling scheme if they transform a Hamiltonian $H$ into $\Hid$ to lowest order, as expressed by the decoupling condition:
\begin{equation}
\frac{1}{m} \sum_{i=0}^{m-1} g_i^\dag H g_i = \frac{1}{D} \Hid.
\label{eq:rec_cond}
\end{equation}
We allow for a possible scaling factor $\frac{1}{D}$ with $D \ge 1$. This factor may require a rescaling of the physical time of the simulated quantum system by $D$. 
Such a quantum simulation 
is only approximate since, in general, the higher orders of $\Havg$ do not vanish. 
However, as $\Havg^{(k)} \sim (\Delta t)^k$, for sufficiently small time delays $\Delta t$ these higher-order corrections become small. 

One of the most important practical questions, which is addressed in the subsequent sections,
is whether, for a specific choice of $H$ and $\Hid$, a decoupling scheme exists at all and how it can be constructed.

\section{Dynamical decoupling schemes for approximate quantum simulation} \label{sec:decoupling_schemes}

In this section, a systematic method for constructing dynamical decoupling schemes on qubit networks is presented
which is capable of transforming the dynamics of a given Hamiltonian $H$ into the dynamics of
a wanted ideal Hamiltonian $\Hid$ by only using Pauli pulses. This way, unwanted environmental interactions can be suppressed
and an ideal Hamiltonian dynamics can be simulated approximately. An extension to a network of qudits of arbitrary dimension $d$ is presented in Sec. \ref{sec:qudits}.

\subsection{Constructing decoupling schemes by solving linear sets of equations} \label{sec:setup}

Using the notation of Sec. \ref{sec:decoupling_basics},
we notice that on the Hilbert space  $\mathcal{H}_\text{sys}$ of an $N$-qubit network,
the set of the $4^N$ operators,
\begin{equation} \label{Sk}
S_j = \sigma_{j_1} \otimes \sigma_{j_2} \otimes \dots \otimes \sigma_{j_N}, \quad j_i \in [0,1,2,3],
\end{equation}
with the base-4 representation $j = j_1 j_2\dots j_N$,
forms a basis for linear operators on $\mathcal{H}_\text{sys}$. Therefore, one
can expand both $N$-qubit system Hamiltonians $H$ and $\Hid$ in this basis with coefficients $\mu_k$ and $\nu_k$, i.e.,
\begin{align} \label{eq:basis_representation}
  H &= \sum_{k=1}^{4^N-1} \mu_k S_k \otimes E_{k} + H_\text{env},\nonumber\\
  \Hid &= \sum_{k=1}^{4^N-1} \nu_k S_k \otimes \id_\text{env} + H_\text{env}.
\end{align}
Here the arbitrary linear operators $E_k$ act on the Hilbert 
space $\mathcal{H}_\text{env}$ of the environment. Furthermore, it is
assumed that we are looking for a dynamical decoupling scheme
which removes all possible couplings between the $N$-qubit system and the environment. 
The part of the Hamiltonians  $H$ and $\Hid$ which acts on the environment only
is denoted $H_\text{env}$
and its precise form is not important for our subsequent discussion.
For the sake of convenience, let us also assume that both $H$ and $\Hid$ are traceless so that
the operator $S_0 \equiv \id$ 
can be omitted from the expansion (\ref{eq:basis_representation}). 

By inserting the basis expansion (\ref{eq:basis_representation}) into the decoupling condition of 
Eq. \eqref{eq:rec_cond}, we obtain the relations
\begin{align}
\frac{D}{m} \sum_{k=1}^{4^N-1} \sum_{j=0}^{m-1} \mu_k \bigl(g_j^\dag S_k g_j \bigr) 
\otimes E_k &= \sum_{k=1}^{4^N-1} 
\nu_k S_k \otimes \id ,\label{eq:dec_inserted}\\ 
\Rightarrow \frac{D}{m} \sum_{k=1}^{4^N-1} \sum_{j=0}^{4^N-1} 
\mu_k c_j \bigl(S_j^\dag S_k S_j \bigr) \otimes E_k &= \sum_{k=1}^{4^N-1} \nu_k S_k  
\otimes \id ,\label{eq:dec_basis} \\
\Rightarrow \frac{D}{m} \sum_{k=1}^{4^N-1} \sum_{j=0}^{4^N-1} \mu_k c_j a_{kj} S_k 
\otimes E_k &= \sum_{k=1}^{4^N-1} \nu_k S_k \otimes \id .
\label{eq:dec_matrix}
\end{align}
Since all of the control pulses $p_j$ are chosen from the set of basis operators $\{S_j\}$, it follows from the basic algebraic properties of the Pauli operators that their products $g_j$ can also be expressed by one of the basis operators, up to a global phase, e.g. $g_j = e^{i\varphi} S_l$. Since this phase vanishes in $\Havgo$, we can replace the sum over the operators $g_j$ in \eqref{eq:dec_inserted} by a sum over the basis operators $S_j$, where we introduce natural number variables $c_j$ which count how often each basis operator $S_j$ occurs in our decoupling scheme $\{g_j\}$. With this replacement, we arrive at relation \eqref{eq:dec_basis}.
Since for all Pauli operators (including the identity operation)
the Clifford-type relation $\sigma_k^\dag \sigma_j^\dag \sigma_k \sigma_j = \pm \id$ holds,
it is also true that $S_k^\dag S_j^\dag S_k S_j = \pm \id$. If we incorporate these signs 
into the variables $a_{kj}$ of unit modulus, we finally arrive at relation  \eqref{eq:dec_matrix}.

Due to the linear independence of the operators $S_k$ in \eqref{Sk}, 
their coefficients can be compared individually in Eq. \eqref{eq:dec_matrix}. This comparison yields
a system of, at most, $4^N-1$ linear equations for the $4^N$ unknown natural numbers $c_j$.
However, we immediately notice a restriction concerning
the solvability of this linear system of equations: 
If for any $k \in [1,4^N-1]$ either $\mu_k = 0$ or $E_k \neq \id$, then the corresponding
expansion coefficient of $\Hid$ has to vanish, i.e., $\nu_k = 0$. This reflects the fact that
any term not present in the originally given 
Hamiltonian $H$ cannot be created by our decoupling scheme in the ideal Hamiltonian $\Hid$.
Furthermore, any operator $S_k$ of the original Hamiltonian $H$ which has nontrivial couplings with the environment, i.e., $E_k \neq \id$, can only be suppressed completely and thus cannot appear in the ideal Hamiltonian $\Hid$.

We obtain the following set of linear equations:
\begin{align}
\frac{D}{m} \sum_{j=0}^{4^N-1} a_{kj} c_j &= \frac{\nu_k}{\mu_k}, 
\quad k \in \mathcal{K}, \label{eq:lin_eq} \\
\mathcal{K} &= \{ k \in [1,4^N-1]: \mu_k \neq 0 \}, \nonumber
\end{align}
with $\sK$ denoting the set of indices of operators $S_k$ which are present in the basis expansion of the 
originally given Hamiltonian $H$.
We need to solve for the non-negative natural numbers
$c_j$ as well as for $D$ and $m$ which are not all independent. This is due to the fact that,
since $m$ is the total number of operators in our decoupling scheme, the relation $\sum c_j = m$ must be fulfilled.
Introducing variables $e_j = D {c_j}/{m}$, our system of equations is finally given by
\begin{equation}
\sum_{j=0}^{4^N-1} a_{kj} e_j = \frac{\nu_k}{\mu_k}, \quad k \in \mathcal{K} .\label{eq:lin_eq_e}
\end{equation}
By construction, the $e_j$ fulfill the relation $\sum e_j = D$, 
and $m$ can be found by determining $m$ as the lowest common denominator for the rational numbers ${e_j}/{D}$. 
One could also choose a larger denominator for $m$. However, this would result in a structurally identical scheme that consists of repetitions of the shorter scheme.

\subsection{Existence of solutions} \label{sec:existence}

Let us now address the question of under which conditions the system of linear equations 
\eqref{eq:lin_eq_e} has suitable solutions. 

Let us first analyze the set of all possible real-valued solutions for the quantities $e_j$. 
The system of equations depends on the previously introduced variables $a_{kj} = \pm 1$ of unit modulus,
which can be computed from the algebraic properties of the operators $S_k$ and $S_j$. 
Doing so for all pairs of our operator basis yields a $4^N \times 4^N$ square matrix 
$A^{(N)} = \{a_{kj}\}$ with entries $\pm 1$. For $N=1$, we can calculate $A^{(1)}$ 
directly from the Pauli operators and obtain a so-called Hadamard matrix,
\begin{equation}
A^{(1)} = \begin{pmatrix}
1 & 1 & 1 & 1 \\
1 & 1 & -1 & -1 \\
1 & -1 & 1 & -1 \\
1 & -1 & -1 & 1
\end{pmatrix}.
\end{equation}
Hadamard matrices have the characteristic property that their entries have the values $\pm 1$ and their rows are mutually orthogonal. Due to the way the  operators $S_k$ are constructed as tensor products of Pauli operators, it follows that higher-order matrices $A^{(N)}$ 
can be constructed from lower-order ones by the recursive relation
\begin{equation}
A^{(N)} = A^{(1)} \otimes A^{(N-1)}.
\end{equation}
This recursive construction is similar to the well-known Sylvester construction for 
Hadamard matrices \cite{syl1867}. From this construction, we can conclude by induction that
$A^{(N)}$ is a Hadamard matrix. A connection between Hadamard matrices and quantum simulation schemes on qubit networks was already discovered in \cite{leu02}, albeit in a more limited context.
Proofs of the stated properties of the system matrix can be found in Appendix \ref{app:matrix} for the more general qudit case.

Since Hadamard matrices have full rank and we use only $\dim \sK < 4^N$ 
rows from the matrix in our system of equations, we can conclude that the system has infinitely many real-valued solutions 
for the variables $e_j$. Now, let $A^{(N)}_\sK$ be the $\dim \sK \times 4^N$ matrix 
resulting from the matrix $A^{(N)}$ 
by including only the rows $\vec{A}^{(N)}_k$ with $k \in \sK$. Then we can express our system of equations \eqref{eq:lin_eq_e}
in compact vector form:
\begin{equation}
A^{(N)}_\sK \cdot \vec{e} = \vec{r}, \qquad \vec{r} = \left(\frac{\nu_k}{\mu_k} \right)_{k \in \sK}.
\end{equation}
Here, $\vec{e}$ is the $4^N$-dimensional vector of our variables $e_j$, and $\vec{r}$ is the $\dim \sK$-dimensional vector of the right-hand side of \eqref{eq:lin_eq_e}.
In order to find the general solution of this linear system of equations, we
first solve for the homogeneous part, $A^{(N)}_\sK \cdot \vec{e} = 0$. 
The linear space of these solutions
has dimension $4^N - \dim \sK$. As
the rows of the Hadamard matrix $A^{(N)}$ are orthogonal and the scalar product of any two 
row vectors $\vec{A}^{(N)}_j$ and $\vec{A}^{(N)}_k$ fulfills the relation
\begin{equation}
\vec{A}^{(N)}_j \cdot \vec{A}^{(N)}_k = 4^N \delta_{jk}, \label{eq:scalar_product}
\end{equation}
we conclude that any multiple of a row of $A^{(N)}$ not contained in $A^{(N)}_\sK$ 
is a solution of the homogeneous equation. Therefore,
the most general homogeneous solution $\vec{e}_0$ is of the form
\begin{equation}
\vec{e}_0 = \sum_{k \not\in \sK} \gamma_k (\vec{A}^{(N)}_k)^T, \quad \gamma_k \in \mathbb{R}.
\end{equation}
A particular solution of the inhomogeneous equation can be constructed by noticing that from
relation \eqref{eq:scalar_product}, one can conclude that 
\begin{equation}
A^{(N)}_\sK \cdot \left(A^{(N)}_\sK \right)^T = 4^N \id_{\dim \sK} .
\end{equation}
This yields the particular solution
\begin{equation}
\vec{e}_r = \frac{1}{4^N} \left(A^{(N)}_\sK \right)^T \cdot \vec{r} .
\end{equation}
Therefore, the most general solution of the system of equations  is given by 
\begin{equation}
\vec{e} = \vec{e}_0 + \vec{e}_r . \label{eq:general_sol}
\end{equation}

Starting from this result, we now need to construct a set of $c_j$ with non-negative integer values. Let us address the issue of non-negativity first.
In order for $c_j$ to be non-negative the quantities $e_j$ must be non-negative, too. 
This latter requirement can be met by starting from an arbitrary solution of the system of equations $\vec{e}$. Because the first row of the Hadamard matrix $\vec{A}^{(N)}_0 = ( 1 1 1 1 \dots )$ is never a part of $A^{(N)}_\sK$ (due to the Hamiltonians being traceless), it is a solution to the homogeneous equation, and so an arbitrary multiple $\gamma_0$ of this first row can always be added to $\vec{e}$. 
In particular, one can choose the multiple as $\gamma_0 = - \min \{e_j\}$ over all entries in a given $\vec{e}$. Adding $\gamma_0 \vec{A}^{(N)}_0$ to $\vec{e}$ ensures that all entries in the resulting solution are non-negative.

Starting from such a non-negative solution, the quantities $c_j$ 
will be integral if the numbers ${e_j}/{D}$ are rational. 
Whether such solutions exist depends entirely on the particular solution $\vec{e}_r$ 
and therefore on the structure of the vector $\vec{r}$. For this purpose,
it is required that all entries of $\vec{r}$ are rational or that they share
at most a common real multiplier so that $\vec{r} = d \vec{r}_0$, with $\vec{r}_0$ 
denoting a rational vector. However, for practical purposes, we can relax this rather stringent condition. Even if the entries of $\vec{r}$ are not rational, it is quite acceptable from a practical point of view to round these quantities to suitable rational numbers. This is possible because dynamical decoupling is already approximate in nature and the lowest order $\Havgo$ is linear in the quantities $c_j$. Therefore, any error originating from rounding will affect $\Havgo$ linearly only, and we conclude that we can always find solutions $c_j$ suitable for decoupling from the linear system \eqref{eq:lin_eq_e}, at least approximately.

\subsection{Practical considerations} \label{sec:limits}

The presented system of linear equations allows us to calculate solutions for the variables $c_j$
 which determine how often each of the basis operators $S_k$ 
appears in the constructed decoupling scheme.
Thus, these variables describe our decoupling scheme completely. 
This means that we can always find a decoupling scheme for any given original Hamiltonian $H$
provided the ideal Hamiltonian $\Hid$ does not contain operators $S_k$ which are either missing in $H$ or which are coupled 
with the environment.

However, the linear system has infinitely many solutions, and consequently infinitely many decoupling schemes exist. Although they all simulate the Hamiltonian $\Hid$ to lowest order, they can differ significantly in their size $m$ and choice of decoupling operators $g_j$ as well as the scaling constant $D$. In general, it is not apparent from these characteristic quantities how a decoupling scheme performs in practice. While their effect on the lowest order of the average Hamiltonian $\Havg$ is identical, the effect on higher orders can be very different. For practical purposes, it is usually preferable to find a decoupling sequence which is short (small $m$) and has a scaling $D$ as close to 1 as possible.

The most straightforward construction of a decoupling scheme is based on the particular solution $\vec{e}_r$, which can be readily calculated and then modified, as described, to yield a positive solution for the $c_j$. Unfortunately, it turns out that the particular solution often produces very large decoupling schemes with large scaling factors $D$. To improve the generated decoupling scheme, we need to exploit the freedom presented by the general homogeneous solution. Unfortunately, it is not apparent how to modify the particular solution in such a way that the resulting scheme has minimal $m$ and $D$.

There is a way to find solutions to the linear system which are guaranteed to have minimal scaling $D$. This is done by employing linear programming. Linear programming is a technique to optimize a linear objective function of a set of variables under certain linear equality and inequality constraints. In our case, we can use linear programming for the set of variables $e_j$ and minimize $D=\sum e_j$ subject to the linear equality constraints \eqref{eq:lin_eq_e}, which will return a solution that is guaranteed to have minimal scaling. However, this approach does not guarantee a minimal scheme size $m$. For small qubit systems with $N\le 5$ we found empirically that the particular solutions generated by typical linear programming solvers often produce sufficiently short decoupling schemes, which are practically usable, but in some instances we were able to manually construct shorter schemes with the same minimal scaling $D$.

Given that the number of variables and the set of equations grow exponentially with the number of qubits $N$, constructing solutions by any method will become increasingly difficult with growing $N$. To find schemes for larger numbers of qubits, it is often better to calculate a solution for a smaller problem instance and then induce a scheme for the full system from the smaller solution.

\section{Example applications} \label{sec:examples}
In the following, we present some basic examples of how to apply our scheme construction to specific scenarios.

\subsection{Protecting a two-qubit interaction from environmentally induced decoherence} \label{sec:twoqubit_gates}

\begin{figure}[b]
\includegraphics[scale=0.8]{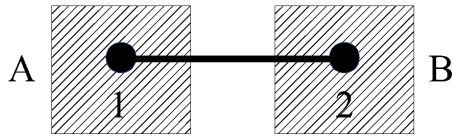} 
\caption{Two interacting qubits $a$ and $b$ coupled to separate environments $A$ and $B$.\label{fig:twoqubits}}
\end{figure}

In this example, we consider two physically separated, distinguishable qubits $1$ and $2$, which interact according to a time-independent interaction Hamiltonian $\Hint$. Furthermore, both qubits are coupled to independent environments $\mathcal{A}$ and $\mathcal{B}$, 
which introduce decoherence and damping. See Fig. \ref{fig:twoqubits} for a visual depiction.
The full dynamics of the system is then described by the most general Hamiltonian,
\begin{align}
H &= \Hint + \Herr , \notag \\
\Hint &= \sum_{i,j=1}^3 h_{ij} \sigma_i^{(1)}  \sigma_j^{(2)} , \notag \\
\Herr &= \sum_{i=1}^3 \sigma_i^{(1)} \otimes A_i + \sum_{i=1}^3 \sigma_i^{(2)} \otimes B_i ,
\end{align}
where $A_i$ and $B_i$ are arbitrary Hermitian operators on their corresponding environments, and $\sigma_i^{(k)}$ means the $i$-th Pauli operator acting on the $k$-th qubit. We ignore potential interactions within and between the environments as they are not relevant to our discussion.

Our goal is to find a decoupling scheme to turn the acting Hamiltonian $H$ into the ideal Hamiltonian $\Hid = \Hint$. For this purpose, we have to solve the system of linear equations (\ref{eq:dec_matrix}). 
The system matrix $A^{(2)} = A^{(1)} \otimes A^{(1)}$ is known and independent of the two Hamiltonians involved. For the coefficients of the vector $\vec{r}$, which determine the inhomogeneous part of the set of equations,
we find
\begin{equation}
\frac{\nu_k}{\mu_k} = \begin{cases}
  1 & \text{ if } S_k \text{ acts on both qubits,} \\
  0 & \text{ if } S_k \text{ acts only on one of the qubits.}
\end{cases}
\end{equation}

The particular solution $\vec{e}_r$ to the set of equations
 allows us to construct the following decoupling scheme by means of the previously described procedure:
\begin{equation} \label{eq:tq_scheme}
\begin{array}{cccccccccccc}
0 & 0 & 0 & 1 & 1 & 1 & 2 & 2 & 2 & 3 & 3 & 3 \\
0 & 0 & 0 & 1 & 2 & 3 & 1 & 2 & 3 & 1 & 2 & 3 .
\end{array}
\end{equation}
Here, the two numbers $i$ and $j$ of each column represent an operator 
$g_k = \sigma_i \otimes \sigma_j$ of our decoupling scheme. 
We see that the scheme consists of a total sequence of $m=12$ pulses.
They involve
free evolution 
(identity pulses applied three times) and a sequence in which all combinations of Pauli 
operators appear exactly once. 
The scaling constant for this decoupling scheme is $D=3$. 
This is the minimal possible value of $D$ which can be verified by 
linear programming.

\subsection{Protecting a $\sqrt{\text{SWAP}}$ gate implementation} \label{sec:swap}
Let us now consider a specific interaction Hamiltonian in the form of a two-qubit Heisenberg Hamiltonian,
\begin{equation} \Hint = \sum_{i=1}^3 \sigma_i^{(1)} \sigma_i^{(2)}. \label{eq:Hint_Heisenberg} \end{equation}
This Hamiltonian can be used to implement the entangling gate $\sqrt{\text{SWAP}}$ if applied over a time interval of duration
$\tau = \frac{\pi}{8}$ (see, e.g., \cite{ss03}). Entangling gates are particularly interesting in quantum information because they can create an entangled state from a
separable two-qubit state. 

For the interaction with the environment, let us now assume that each qubit is coupled
to a harmonic 
oscillator as described by the Hamiltonian
\begin{equation} \Herr = \lambda ( \sigma_+^{(1)} a + \sigma_-^{(1)} a^\dag + \sigma_+^{(2)} b + 
\sigma_-^{(2)} b^\dag ). \label{eq:Herr_Heisenberg} 
\end{equation} 
Here, $a$, $a^\dag$ and $b$, $b^\dag$ are the creation and annihilation operators 
of the two oscillators, and $\sigma_\pm = \frac{1}{2} (\sigma_1 \pm i \sigma_2)$. $\lambda$ characterizes
the common strength of the coupling. 

The general two-qubit protection scheme \eqref{eq:tq_scheme} can be used to protect this $\sqrt{\text{SWAP}}$ gate implementation.
However, the Hamiltonian in \eqref{eq:Hint_Heisenberg}
contains only three out of nine possible two-qubit basis operators and 
the error terms in \eqref{eq:Herr_Heisenberg} involve only four out of six possible operators. 
Taking these special circumstances into account, 
we can simplify the system of equations (\ref{eq:dec_matrix}) 
by
omitting eight equations. As a result, we find the significantly
simpler dynamical decoupling scheme:
\begin{equation} \label{eq:swap_scheme}
\begin{array}{cccc}
0 & 1 & 3 & 2 \\
0 & 1 & 3 & 2.
\end{array}
\end{equation}
This particular protection scheme involves only four 
decoupling operators and has an improved scaling factor of $D=1$. This means that no rescaling of the interaction time is necessary, therefore allowing the gate to be implemented faster. Furthermore, the control operators $g_k$ required for the special scheme
can be implemented with the help of pulses $p_k$ utilizing only $\sigma_1$ and $\sigma_2$ pulses, 
whereas the general scheme also requires $\sigma_3$ pulses, so
it should be easier to implement experimentally.  

To study the effectiveness of our protection schemes, we conducted numerical simulations of the achievable state fidelity depending on the error strength $\lambda$ and the pulse delay time $\Delta t$. The results are very encouraging because even under the influence of strong errors, a state fidelity close to $1$ is reached with pulse delays as large as $\Delta t = \frac{\tau}{4}$. A detailed presentation of the numerical results and a comparison between the two decoupling schemes is included in Appendix \ref{sec:numerical}.

\subsection{Removing diagonal couplings in a closed $4$-qubit chain}
\begin{figure}[b]
\includegraphics[scale=0.8]{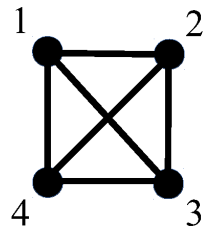} 
\caption{A quadratic closed chain of four qubits with diagonal couplings. \label{fig:diagonal}}
\end{figure}

As another example, let us consider a closed chain of four qubits with $\sigma_3 \otimes \sigma_3$ coupling between each qubit pair, as depicted in Fig. \ref{fig:diagonal}. We can describe the couplings by the Hamiltonian
\begin{equation}
H = \sum_{\substack{i,j=1 \\ i < j}}^4 h_{i,j} \sigma_3^{(i)} \sigma_3^{(j)}.
\end{equation}
Our goal is to eliminate the diagonal couplings between qubits 2 and 4 and between qubits 1 and 3. The ideal Hamiltonian without these diagonal couplings can be written as
\begin{equation}
\Hid = \sum_{i=1}^4 h_{i,i+1} \sigma_3^{(i)} \sigma_3^{(i+1)},
\end{equation}
if we assume that the indexes wrap around, i.e., qubit 5 is just qubit 1 again. This scenario is an example of selective decoupling where we want to remove only certain parts of the given Hamiltonian.

Setting up the linear system is straight-forward. The system matrix is $A^{(4)} = A^{(1)} \otimes A^{(1)} \otimes A^{(1)} \otimes A^{(1)}$. Our set $\mathcal{K}$ of basis operator indices occurring in $H$ consists of $\{3300, 0330, 0033, 3003, 3030, 0303\}$, with the numbers in base-4 notation so that a number corresponds to an operator $abcd \Rightarrow \sigma_a \otimes \sigma_b \otimes \sigma_c \otimes \sigma_d$. Finally, we set the vector $\vec{r} = (1,1,1,1,0,0)$, meaning that we want to keep the first four operators in $\mathcal{K}$ and eliminate the other two.

We employ linear programming to construct a decoupling scheme for this scenario. The solution given by our linear programming solver leads to the following decoupling scheme:
\begin{equation}
\begin{array}{cccc}
0 & 0 & 1 & 0 \\
0 & 0 & 1 & 1 \\
0 & 0 & 0 & 1 \\
0 & 0 & 0 & 0.
\end{array}
\end{equation}
The scaling factor of this scheme is $D=2$, meaning that the interaction time has to be doubled to compensate for the reduced interaction strength.

\subsection{Modifying individual interaction strengths}
\begin{figure}[b]
\includegraphics[scale=0.8]{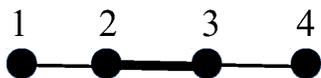} 
\caption{A linear chain of four qubits with nearest-neighbour interactions, where the inner coupling is twice as strong as the two outer couplings. \label{fig:chain}}
\end{figure}
As a final example, we demonstrate how to construct schemes which can modify individual coupling strengths between qubits. Consider a linear chain of four qubits with nearest-neighbor $XX$ interactions which are all equally strong. The Hamiltonian describing these interactions is 
\begin{equation}
H = \sum_{i=1}^3 J ( \sigma_1^{(i)} \sigma_1^{(i+1)} + \sigma_2^{(i)} \sigma_2^{(i+1)} ),
\end{equation}
with $J$ an arbitrary coupling strength. Imagine that we would like to reduce the coupling strengths between qubits 1 and 2 and between qubits 3 and 4 by half, as depicted in Fig. \ref{fig:chain}. The resulting Hamiltonian would be
\begin{align}
\Hid =& J (\sigma_1^{(2)} \sigma_1^{(3)} + \sigma_2^{(2)} \sigma_2^{(3)}) \notag \\ &+ \frac{1}{2} J (\sigma_1^{(1)} \sigma_1^{(2)} + \sigma_2^{(1)} \sigma_2^{(2)} + \sigma_1^{(3)} \sigma_1^{(4)} + \sigma_2^{(3)} \sigma_2^{(4)}) .
\end{align}

Our system matrix is $A^{(4)}$ as before; the relevant operator indices in our set $\mathcal{K}$ are $\{1100, 2200, 0110, 0220, 0011, 0022\}$ with corresponding entries in the vector $\vec{r} = (0.5, 0.5, 1.0, 1.0, 0.5, 0.5)$. This way we ensure that the couplings in the middle are kept intact, while the couplings at the outer edges are reduced.

By employing linear programming again, we find the following decoupling scheme with a scaling $D=1$:
\begin{equation}
\begin{array}{cccc}
0 & 0 & 0 & 0 \\
0 & 0 & 0 & 3 \\
0 & 0 & 0 & 3 \\
0 & 0 & 0 & 0.
\end{array}
\end{equation}
With a slight modification to this example one can simulate the perfect state transfer Hamiltonian on a linear chain of qubits discovered by M. Christandl {\it et al.} \cite{cdel04} and independently by G. Nikolopoulos {\it et al.} \cite{npl04}.

\section{Approximate quantum simulation in qudit systems} \label{sec:qudits}
In this section, we briefly discuss how the construction of dynamical decoupling schemes for approximate
quantum simulation can be generalized to qudit systems. Technical
details of the necessary generalizations are presented in the appendices.

A qudit is a finite-dimensional quantum system with a $d$-dimensional 
Hilbert space $\mathbb{C}^d$. Correspondingly, 
we replace the standard Pauli operators for qubits 
by a generalized set of spin operators \cite{pr04},
\begin{equation}
\sigma_{j,k} = \sum_{l=0}^{d-1} \omega^{jl} \ket{l} \bra{l+k}, \quad j,k \in [0,d-1] \label{eq:spin_ops},
\end{equation}
with $\omega = e^{2\pi i / d}$ and $\ket{l} \equiv \ket{l \bmod d}$. 
For $d=2$, these operators are equivalent to the Pauli operators up to phase factors.
For arbitrary values of $d>2$, these operators are still unitary but no longer
Hermitian. They obey the following characteristic algebraic property as derived in Appendix \ref{app:spin_ops}:
\begin{equation}
\sigma_{s,t}^\dag \sigma_{j,k}^\dag \sigma_{s,t} \sigma_{j,k} = \omega^{jt-ks} \, \id \label{eq:qudits_char}.
\end{equation}

For an $N$-qudit system, the $d^{2N}$ tensor products of these generalized spin operators, i.e.
\begin{equation}
S_{j,k} = \sigma_{j_1,k_1} \otimes \sigma_{j_2,k_2} \otimes \dots \otimes \sigma_{j_{N},k_{N}},
\label{eq:qudits_base_ops}
\end{equation}
with $j=j_1 j_2\dots j_N$ and $k=k_1 k_2\dots k_{N}$ in base-$d$ representation,
 still form a complete operator basis. Therefore,
the same approach as discussed in Sec. \ref{sec:setup}
can be used to set up a linear system of equations derived from the decoupling condition to lowest order. 
Analogous to the steps involved in
Eqs. \eqref{eq:basis_representation} to \eqref{eq:dec_matrix},
the Hamiltonians $H$ and $\Hid$ can be expanded in the basis of these operators 
$S_{j,k}$, and
the variables $c_{j,k}$, which denote how often each 
operator $S_{j,k}$ occurs in the decoupling scheme, can be determined from the resulting 
system of linear equations.

From Eq. \eqref{eq:qudits_char},
we notice that now the corresponding matrix $A$ 
has complex entries involving the $d$-th unit roots $\omega^n$, $n \in [0,d-1]$. 
By constructing $A$ analogously to the case of qubit systems,
we note that its first row still consists of unit entries,
while all other rows contain each of the unit roots equally often.
In addition, all rows are linearly independent. 
Therefore, $A$ is now a complex-valued Hadamard matrix \cite{b62}. Details of the properties of $A$ with proofs are given in Appendix \ref{app:matrix}.

As our system of equations is now complex
but the variables $c_{j,k}$ (or their replacements $e_{j,k}$) still have to be real valued, it is not immediately apparent whether a (suitable) solution still exists. A detailed analysis in Appendix \ref{app:eq} reveals that it is, indeed, still possible to find suitable solutions subject to the same conditions that apply to the qubit case.
Thus we conclude that with these generalizations,
our method works just as well for qudit networks.

\section{Conclusions and outlook}
We have presented a general
method for constructing dynamical decoupling schemes for networks of qudit systems. This
dynamical decoupling method is capable 
of simulating an ideal Hamiltonian of a general qudit network
and
of decoupling this system from unwanted environmental influences approximately.
The proposed decoupling schemes 
involve local applications
of Pauli operators or of their higher-dimensional generalizations only, so that they
offer interesting perspectives for experimental applications. 

We have exemplified the usage of our method in three different scenarios. Those scenarios cover typical cases for our method: decoupling a system from its environment, selectively decoupling particular interactions between qubits, and modifying individual coupling strengths. Although they were presented in separate examples, our method also allows for any combination of these cases in a single scenario.

Due to the exponential scaling with the number of qubits $N$ of the linear system used in our method and the infinitely many solutions that exist, there are some practical limitations to be aware of. Although it is straight-forward to calculate a scheme from the particular solution, these schemes are typically very large and therefore not practical. To improve the generated scheme, one can, in principle, make use of the degrees of freedom presented by the general homogeneous solution, but so far no systematic procedure is known which guarantees a shorter and superior scheme. We briefly discussed the use of linear programming as a feasible workaround to find more suitable solutions and made use of it to find some of the schemes presented in the examples.

\begin{acknowledgments}Financial support by the BMBF-project QuOReP and by CASED is acknowledged. PB acknowledges support by RVO Grant No. 68407700. 
\end{acknowledgments}
\appendix

\section{Basic spin operator properties} \label{app:spin_ops}
We show some algebraic properties of the generalized spin operators $\sigma_{j,k}$ given in Eq. \eqref{eq:spin_ops}.

The adjoints of these operators are given by
\begin{align}
\sigma_{j,k}^\dag &= \sum_{l=0}^{d-1} (\omega^*)^{jl} \ket{l+k} \bra{l} = \sum_{l=0}^{d-1} \omega^{(d-j)l} \ket{l+k} \bra{l} \notag \\
&= \omega^{jk} \sum_{m=k}^{d+k-1} \omega^{(d-j)m} \ket{m} \bra{m+(d-k)} \notag \\
&= \omega^{jk} \sigma_{d-j,d-k} .
\end{align}

The product of two spin operators yields
\begin{align}
\sigma_{j,k} \sigma_{s,t} &= \sum_{l=0}^{d-1} \omega^{jl+s(l+k)} \ket{l} \bra{l+k+t} \notag \\
&= \omega^{sk} \sigma_{j+s,k+t} .
\end{align}
The generalized spin operators are unitary because
\begin{equation}
\sigma_{j,k} \sigma_{j,k}^\dag = \omega^{jk} \sigma_{j,k} \sigma_{d-j,d-k} = \omega^{jk} \omega^{(d-j)k} \sigma_{0,0} = \id .
\end{equation}

Finally, we obtain the characteristic relation
\begin{align}
\sigma_{j,k}^\dag \sigma_{s,t} \sigma_{j,k} &= \omega^{jk} \omega^{tj} \omega^{(s+j)(d-k)} \sigma_{d+s,d+t} \notag \\
&= \omega^{jt-ks} \sigma_{s,t} . \label{eq:A_entries}
\end{align}

\section{Properties of the system matrix $A$} \label{app:matrix}
The system matrix $A$ is a $d^{2N} \times d^{2N}$ complex matrix whose entries 
$a_{(s,t),(j,k)}$ are determined by the base operators $S_{j,k}$ and $S_{s,t}$ given in Eq. \eqref{eq:qudits_base_ops}, 
according to $S_{j,k}^\dag S_{s,t} S_{j,k} = a_{(s,t),(j,k)} S_{s,t}$.
The adjoint of $S_{j,k}$ is given by
\begin{align}
S_{j,k}^\dag &= \bigotimes_{i=N-1}^0 \omega^{j_{i} k_{i}} \sigma_{d-j_{i},d-k_{i}} \\
&= \omega^{\dotp{j}{k}} S_{d^N-j, d^N-k},
\end{align}
with $\dotp{j}{k} := \sum_{i=0}^{N-1} j_i k_i$ a scalar product of the base-$d$ representations of $j$ and $k$. Then it follows that
\begin{align}
S_{j,k}^\dag S_{s,t} S_{j,k} &= \bigotimes_{i=N-1}^0 \omega^{j_i t_i - k_i s_i} \sigma_{s_i,t_i} \notag \\
&= \omega^{\dotp{j}{t} - \dotp{k}{s}} S_{s,t}  .
\end{align}
The matrix $A$ is Hermitian because
\begin{equation}
a_{(s,t),(j,k)}^* = \left(\omega^{\dotp{j}{t} - \dotp{k}{s}} \right)^* = \omega^{-\dotp{j}{t} + \dotp{k}{s}} = a_{(j,k),(s,t)} .
\end{equation}

The entries of any row vector $\vec{A}_{(s,t)}$ of the system matrix $A$ sum up to zero with the only
exception being the first row $\vec{A}_{(0,0)}$. 
This follows from the fact that for any integral power $k\neq 0$ of the unit root, $\omega^k$, 
it holds that $\sum_{j=0}^{d-1} \omega^{kj} = 0$. 
We assume that $s_n$ is a non-zero component of the base-$d$ representation of $s$.
Thus, we find for the sum over the row $\vec{A}_{(s,t)}$ the expression
\begin{align}
& \sum_{j=0}^{d^N-1} \sum_{k=0}^{d^N-1} a_{(s,t),(j,k)} = \sum_{\substack{j_0,\dots ,j_{N-1}=0 \\ k_0,\dots ,k_{N-1}=0}}^{d-1} \prod_{i=0}^{N-1} \omega^{j_i t_i} \omega^{-k_i s_i} \notag \\
&= \sum_{\substack{j_0,\dots ,j_{N-1}=0 \\ k_0,\dots ,k_{n-1},k_{n+1}, \\ \dots ,k_{N-1}=0}}^{d-1} \prod_{\substack{i=0 \\ i \neq n}}^{N-1} \omega^{j_i t_i} \omega^{-k_i s_i} \omega^{j_n t_n} \sum_{k_n=0}^{d-1} \omega^{-k_n s_n} \notag \\
&= 0 .
\end{align}
The same result is obtained
if any component $t_n$ of the base-$d$ representation of $t$ is non-zero.
 Only if $s=t=0$ does the sum not vanish. In this case, all terms in the sum are equal to $1$
 and performing the sums yields the result $d^{2N}$.

As a direct consequence, we find, for the scalar product between any two row vectors 
$\vec{A}_{(s,t)}$ and $\vec{A}_{(u,v)}$,
\begin{align}
\vec{A}_{(s,t)}\cdot \vec{A}_{(u,v)} &= \sum_{j,k=0}^{d^N-1} \omega^{ \dotp{j}{t} - \dotp{k}{s}} \omega^{-\dotp{j}{v} +\dotp{k}{u}} \notag \\
&= \sum_{j,k=0}^{d^N-1} \omega^{\vec{j}\cdot (\vec{t}-\vec{v}) - \vec{k}\cdot (\vec{s}-\vec{u})} \notag \\
&= d^{2N} \delta_{s,u} \delta_{t,v} , \label{eq:scalar}
\end{align}
which corresponds to the sum over the row $\vec{A}_{(s-u,t-v)}$.

From Eq. \eqref{eq:A_entries}, it follows that for any row vector $\vec{A}_{(s,t)}$, there is 
another row vector $\vec{A}_{(d^N-s,d^N-t)}$ whose entries are the complex conjugate of the former, i.e.,
\begin{align}
a_{(s,t),(j,k)} &= \omega^{\dotp{j}{t}-\dotp{k}{s}} = \left(\omega^{-\dotp{j}{t}+\dotp{k}{s}} \right)^* \notag \\
&= a_{(-s,-t),(j,k)}^* = a_{(d^N-s,d^N-t),(j,k)}^* .
\end{align}

\section{Real-valued solutions of the complex system of linear equations} \label{app:eq}
Consider two Hamiltonians $H$ and $\Hid$ given by 
\begin{align}
H &= \sum_{(j,k) \in \mathcal{K}} \mu_{j,k} S_{j,k} \otimes E_{j,k} + \Henv, \\
\Hid &= \sum_{(j,k) \in \mathcal{K}} \nu_{j,k} S_{j,k} \otimes \id_\text{env} + \Henv, 
\end{align}
where $\mathcal{K}$ is a subset of $[0,d^N-1] \times [0,d^N-1]$ and does not contain the pair $(0,0)$.
Therefore, these Hamiltonians are traceless. 
We want to find real-valued solutions to the complex system of linear equations
\begin{equation} \label{eq:lin_eq_complex}
A_{\mathcal{K}} \cdot \vec{e} = \vec{r}, \qquad \vec{r} = \left( \frac{\nu_{j,k}}{\mu_{j,k}} \right)_{(j,k) \in \mathcal{K}},
\end{equation}
with $A_{\mathcal{K}}$ being the submatrix of $A$ consisting of all the rows
 $\vec{A}_{(j,k)}$ for $(j,k) \in \mathcal{K}$. 
This system of equations
is only solvable if, for any $\mu_{j,k}=0$,
 the corresponding $\nu_{j,k}$ also vanishes. (Compare with the discussion in Sec. \ref{sec:setup}.) 
In the following, it is therefore assumed 
that $\mu_{j,k} \neq 0$ for any $(j,k) \in \mathcal{K}$.

As the Hamiltonians $H$ and $\Hid$ are Hermitian, we can conclude that
$(d^N-j,d^N-k) \in \mathcal{K}$, 
and $\mu_{j,k} = \omega^{\dotp{j}{k}} \mu_{d^N-j,d^N-k}^*$, $\nu_{j,k} = \omega^{\dotp{j}{k}} \nu_{d^N-j,d^N-k}^*$
provided
$(j,k) \in \mathcal{K}$.
Therefore, we obtain the relation
\begin{equation}
\frac{\nu_{j,k}}{\mu_{j,k}} = \left( \frac{\nu_{d^N-j,d^N-k}}{\mu_{d^N-j,d^N-k}} \right)^* .
\end{equation}
With our knowledge of the scalar products of $A$'s row vectors, we can conclude that a 
particular solution to the linear system of equations is given by
\begin{equation}
\vec{e}_r = \frac{1}{d^{2N}} A_\mathcal{K}^\dag \cdot \vec{r},
\end{equation}
since $\frac{1}{d^{2N}} A_\mathcal{K} A_\mathcal{K}^\dag = \id_\mathcal{K}$.
This particular solution is real-valued because for any entry of the vector $\vec{e}_r$, we find
\begin{align}
(\vec{e}_r)_{(s,t)} &= \frac{1}{d^{2N}} \sum_{(j,k)\in \mathcal{K}} a^*_{(s,t),(j,k)} \frac{\nu_{j,k}}{\mu_{j,k}} \notag \\
&= \frac{1}{d^{2N}} \sum_{(j,k)\in \mathcal{K}} \frac{1}{2} \Big[ a^*_{(s,t),(j,k)} \frac{\nu_{j,k}}{\mu_{j,k}} \notag \\
&+ a^*_{(d^N-s,d^N-t),(d^N-j,d^N-k)} \frac{\nu_{d^N-j,d^N-k}}{\mu_{d^N-j,d^N-k}} \Big] \notag \\
&= \frac{1}{2 d^{2N}} \sum_{(j,k)\in \mathcal{K}} \Big[ a^*_{(s,t),(j,k)} \frac{\nu_{(j,k)}}{\mu_{(j,k)}} \notag \\
&+ a_{(s,t),(j,k)} \left(\frac{\nu_{j,k}}{\mu_{j,k}}\right)^* \Big]. \label{eq:realvalued}
\end{align}
The expression in brackets in \eqref{eq:realvalued} 
is of the form $c + c^*$ so that it is real valued. 
Adding
a solution of the homogeneous equation $A_\mathcal{K} \cdot \vec{e}_0 = 0$ to $\vec{e}_r$ yields again a solution
of the linear system of Eq. \eqref{eq:lin_eq_complex}. 
From the scalar products in Eq. \eqref{eq:scalar}, we can conclude that any row vector 
$\vec{A}_{(j,k)}, (j,k) \not \in \mathcal{K}$ is a solution of the homogeneous system of equations.
Therefore, the most general solution of the linear system of equations is given by 
\begin{equation}
\vec{e} = \frac{1}{d^{2N}} A^\dag_\mathcal{K} \cdot \vec{r} + 
\sum_{(j,k) \not \in \mathcal{K}} \alpha_{(j,k)} \vec{A}_{(j,k)} .
\end{equation}
As $(0,0) \not \in \mathcal{K}$, this construction allows us to find at least one real-valued and
non-negative solution of the form $\vec{e} = \vec{e}_r + \alpha \vec{A}_{(0,0)}$ 
with $\alpha$ chosen so that all entries in $\vec{e}$ are non-negative. 
Approximating the real values by rational numbers 
allows us to construct a decoupling scheme from $\vec{e}$, as explained in Sec. \ref{sec:existence}.

\section{Numerical simulations for the $\sqrt{\text{SWAP}}$ gate} \label{sec:numerical}

\begin{figure}[t]
    \includegraphics[scale=1]{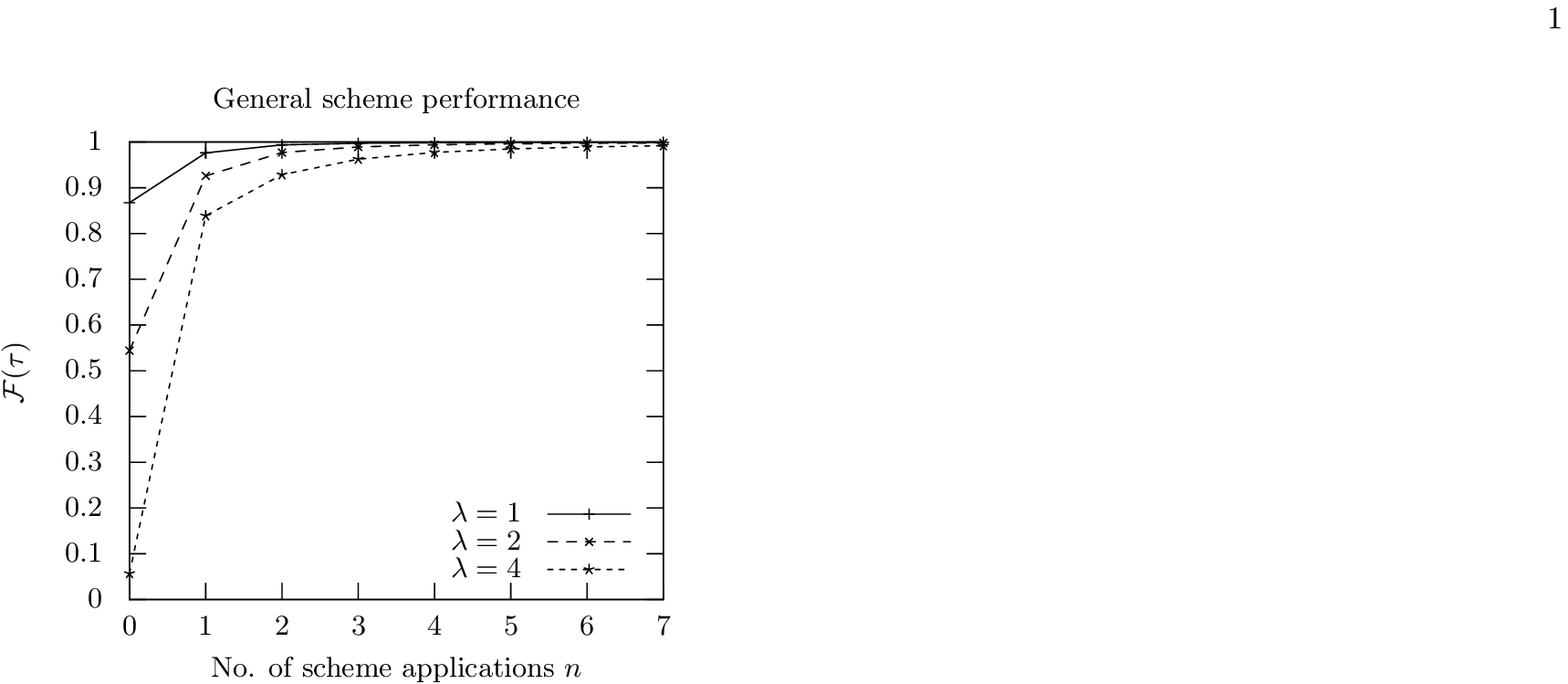}
    \caption{Performance of the general two-qubit interaction scheme for protecting 
the two-qubit gate against different strengths of environmental couplings, with dependence of the fidelity according to Eq. (\ref{fidelity}) on the number of applications
$n$  of the dynamical decoupling scheme;
the scaling $D=3$ of the dynamical decoupling
scheme has been compensated by extending the actual 
interaction time from its ideal value $\tau$ to $3 \tau$.}
    \label{fig:general}
\end{figure}

In order to demonstrate the effectiveness of our method, we have performed numerical simulations for the scenario outlined in Sec. \ref{sec:swap}. We assume that two qubits are initially prepared in a 
separable state $\ket{\Psi}$ and evolve under the influence of the Hamiltonian $H = \Hint + \Herr$ as given by Eqs. \eqref{eq:Hint_Heisenberg} and \eqref{eq:Herr_Heisenberg}.
Without environmental interaction, this state 
would, after a time $\tau=\frac{\pi}{8}$, evolve to the quantum state $\Uid(\tau) \ket{\Psi}$ under the action of the ideal Heisenberg Hamiltonian 
$\Hint$.
The final state resulting from time evolution under the total Hamiltonian
$H$ and under active decoupling controls can be represented in the form
\begin{equation}
U(D \tau) \bigl( \ket{\Psi} \otimes \ket{0}_\text{env} \bigr) = 
\sum_j \alpha_j \ket{\Psi_j} \otimes \ket{e_j}_\text{env},
\end{equation}
with
$\ket{e_j}_\text{env}$ constituting an orthonormal 
basis of the environmental Hilbert space. 
($\ket{0}_\text{env}$ denotes the initially prepared ground state of the environmental harmonic oscillators.) We have taken into account that the interaction time may need to be rescaled by the factor $D$ of the decoupling scheme which is used.
This result can be compared with
the ideal case by means of the reduced state fidelity
\begin{equation}\label{fidelity}
\mathcal{F}(\tau) = \sum_j \bigl\vert \alpha_j \bigr\vert^2  \bigl\vert 
\bra{\Psi} \Uid^\dag(\tau) \ket{\Psi_j} \bigr\vert^2
\end{equation}
in a convenient way.

In Fig. \ref{fig:general}, the results of numerical
simulations of the general two-qubit protection scheme \eqref{eq:tq_scheme} are represented for different choices of the coupling strengths $\lambda$. 
The graph shows the final fidelity $\mathcal{F}(\tau)$ after the gate implementation time $\tau$ 
and its dependence on the number of times the decoupling scheme has been applied. 
The pulse sequences of the dynamical decoupling scheme
are distributed equally over the whole interaction time $D\tau$.
Therefore, in order to apply the decoupling scheme $n$ times 
it is necessary to implement a control pulse frequency of magnitude
\begin{equation}
\frac{1}{\Delta t} = \frac{m \cdot n}{D \tau},
\end{equation}
with $m=12$ and $D=3$ for the general two-qubit protection scheme of Sec. \ref{sec:twoqubit_gates}. 
It is apparent that
with increasing values of $n$ and consequently smaller times between subsequent control pulses
$\Delta t$, the performance of the decoupling procedure increases. 
This is expected as the higher-order terms $\Havg^{(k)}$ in the average Hamiltonian 
are of the order of $(\Delta t)^k$. But even if the scheme is applied only once or twice,
 the increase of the fidelity is noticeable, particularly in the presence of stronger couplings
to the environment.

\begin{figure}[t]
    \includegraphics[scale=1]{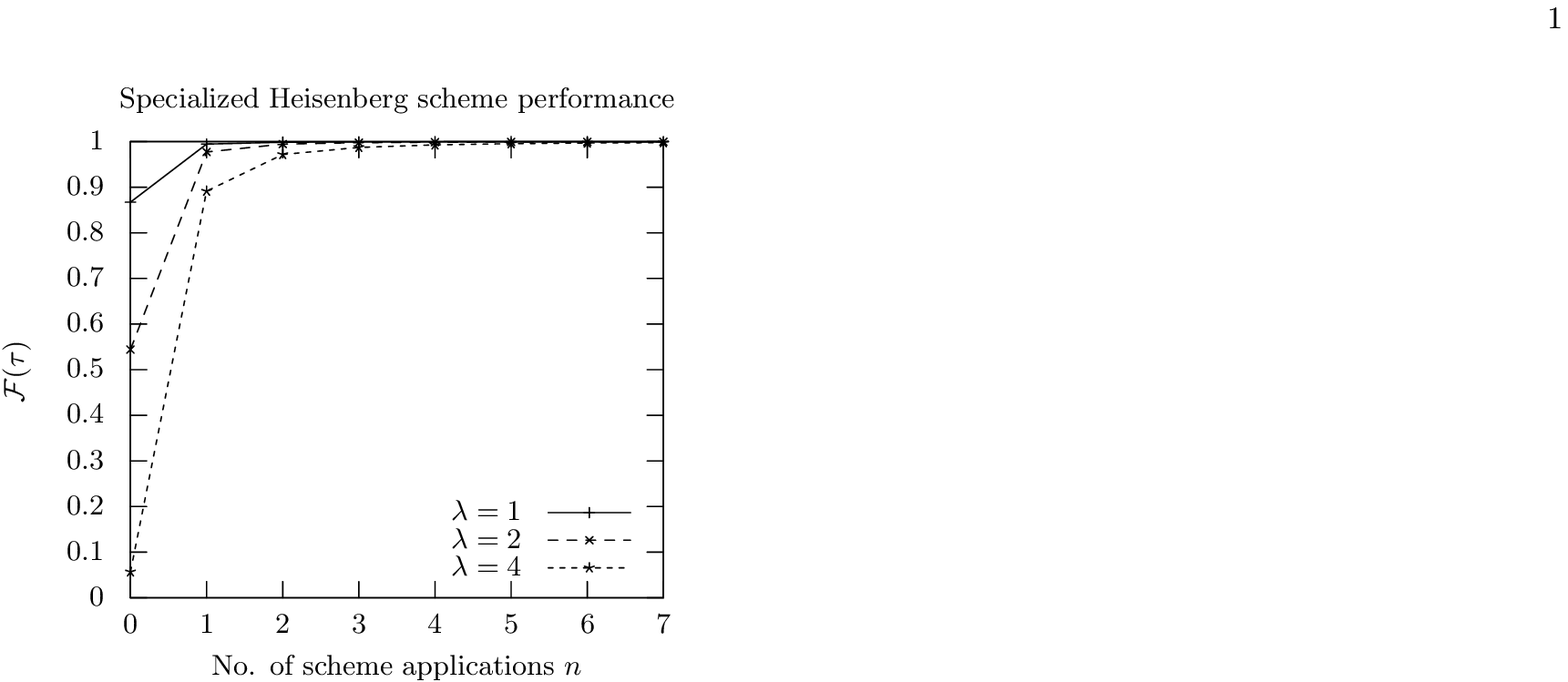}
    \caption{Performance of the special Heisenberg protection scheme \eqref{eq:swap_scheme}: The fidelity of the system after 
the gate interaction time $\tau$ and its dependence on the number $n$ of applications of the dynamical decoupling scheme are plotted.}
    \label{fig:heisenberg}
\end{figure}
Figure \ref{fig:heisenberg} presents results from the same numerical 
simulations, but this time employing the specialized decoupling scheme \eqref{eq:swap_scheme}.
Compared to the performance of the general two-qubit scheme, the fidelity is slightly improved. This is an additional practical advantage of the specialized scheme over the general one, besides those mentioned in Sec. \ref{sec:swap}.

\end{document}